\documentclass[a4paper,10pt,twocolumn]{article}
\pdfoutput=1
\usepackage{graphicx}
\usepackage{grffile}
\usepackage{xcolor}
\usepackage{url}
\usepackage[colorlinks=true,linkcolor=blue,citecolor=blue]{hyperref}
\usepackage[expproduct=cdot]{siunitx} 

\usepackage{relsize} 
\usepackage{amsmath}
\usepackage{amssymb}
\usepackage[utopia,greekuppercase=italicized]{mathdesign}\edef\partial{\mathchar\number\partial\noexpand\!} 

\usepackage[T1]{fontenc} 

\usepackage{ctable} 
\usepackage{tabularx} 

\usepackage{cite} 

\definecolor{royalblue4}{HTML}{27408B}
\definecolor{red4}{HTML}{8B0000}
\definecolor{green4}{HTML}{008b00} 

\usepackage{dblfloatfix} 
\usepackage{fixltx2e} 

\usepackage[font=footnotesize,labelfont=bf,labelsep=endash,margin=0pt]{caption} 
\usepackage[title,header]{appendix}

\newlength{\myleftmargin} 
\newlength{\mytopmargin} 
\newlength{\myrightmargin} 
\newlength{\mybottommargin} 
\usepackage[runin]{abstract}

\let\paragraphold\paragraph
\renewcommand*{\paragraph}[1]{\paragraphold{#1.}} 
\newcommand{\keywords}[1]{\vspace{2mm}\noindent\textbf{Key words:} #1} 
\newcommand{\pagewidetitle}[3] 
{%
    \twocolumn%
        [%
            \vskip-5mm%
            \begin{@twocolumnfalse}%
                #1%
                #2%
                \vspace{5mm}%
            \end{@twocolumnfalse}%
        ]%
        #3%
}

\newlength{\figurewidth}

\definecolor{pbuenzli}{HTML}{8B008B}
\makeatletter
\long\def\tlist@if@empty@nTF #1{%
\expandafter\ifx\expandafter\\\detokenize{#1}\\%
\expandafter\@firstoftwo
\else
\expandafter\@secondoftwo
\fi
}

\newcommand*\pbuenzli[2][]{
{\color{pbuenzli}\it #2}%
\tlist@if@empty@nTF{#1}{}{\footnote{\color{pbuenzli}\it #1}}
}
\makeatother

\newcommand{\ie}{{\it i.e.}}

\newcommand{\eg}{{\it e.g.}}
\newcommand{\etal}{{\it et\ al.}}

\newcommand{\der}{\mathrm{d}}
\newcommand{\p}{\partial}
\newcommand{\pd}[2]{\frac{\partial #1}{\partial #2}}
\newcommand{\tpd}[2]{\tfrac{\partial #1}{\partial #2}}
\newcommand{\td}[2]{\tfrac{\der #1}{\der #2}}
\newcommand{\fd}[2]{\frac{\der #1}{\der #2}}
\newcommand{\Order}{\mathrm{O}}

\newcommand{\e}{\mathrm{e}}

\renewcommand{\b}[1]{{\boldsymbol{#1}}} 

\newcommand{\da}{\ensuremath{\text{day}}} 
\newcommand{\days}{\ensuremath{\text{days}}} 

\newcommand{\um}{\ensuremath{\micro\metre}}

\usepackage{relsize} 
\newcommand{\bmu}{\text{BMU}} 

\newcommand{\ob}{\text{\textsmaller{O}b}}

\newcommand{\ocy}{\text{\textsmaller{O}cy}}

\newcommand{\wnt}{\text{\textsmaller{W}nt}}

\newcommand{\rankl}{\text{\textsmaller{RANKL}}}

\newcommand{\tgfb}{\text{\textsmaller{TGF$\beta$}}}

\newcommand{\kform}{\text{$k_\text{form}$}} 

\newcommand{\dburial}{\text{$D_\text{burial}$}} 
\newcommand{\fracocyob}{\text{$\nu^\ocy_\ob$}}
\newcommand{\fracocylac}{\text{$\nu^\ocy_\text{lac}$}}

\begin{document}
\title{\bf Osteocytes as a record of bone formation dynamics: \\A mathematical model of osteocyte generation in bone matrix}
\author{P R Buenzli$^\text{a,1}$}

\date{\small \vspace{-2mm}$^\text{a}$School of Mathematical Sciences, Monash University, Clayton, VIC3800, Australia\\\vskip 1mm \normalsize \today\vspace*{-5mm}}

\pagewidetitle{
\maketitle
}{
\begin{abstract}
The formation of new bone involves both the deposition of bone matrix, and the formation of a network of cells embedded within the bone matrix, called osteocytes. Osteocytes derive from bone-synthesising cells (osteoblasts) that become buried in bone matrix during bone deposition. The generation of osteocytes is a complex process that remains incompletely understood. Whilst osteoblast burial determines the density of osteocytes, the expanding network of osteocytes regulates in turn osteoblast activity and osteoblast burial. In this paper, a spatiotemporal continuous model is proposed to investigate the osteoblast-to-osteocyte transition.  The aims of the model are (i) to link dynamic properties of osteocyte generation with properties of the osteocyte network imprinted in bone, and (ii) to investigate Marotti's hypothesis that osteocytes prompt the burial of osteoblasts when they become covered with sufficient bone matrix. Osteocyte density is assumed in the model to be generated at the moving bone surface by a combination of osteoblast density, matrix secretory rate, rate of entrapment, and curvature of the bone substrate, but is found to be determined solely by the ratio of the instantaneous burial rate and matrix secretory rate. Osteocyte density does not explicitly depend on osteoblast density nor curvature. Osteocyte apoptosis is also included to distinguish between the density of osteocyte lacuna and the density of live osteocytes. Experimental measurements of osteocyte lacuna densities are used to estimate the rate of burial of osteoblasts in bone matrix. These results suggest that: (i) burial rate decreases during osteonal infilling, and (ii) the control of osteoblast burial by osteocytes is likely to emanate as a collective signal from a large group of osteocytes, rather than from the osteocytes closest to the bone deposition front.

    \keywords{bone formation, osteocyte, osteoblast burial, matrix synthesis}
\end{abstract}
}{
\protect\footnotetext[1]{Corresponding author. Email address: \texttt{pascal.buenzli@monash.edu}}
}

\section{Introduction}
Bone is an adaptive biomaterial that is able to detect and remove micro-damage, and that optimises its shape and microstructure to the mechanical loads it carries~\cite{martin-burr-sharkey,cowin-handbook}. These properties of bone are conferred by a network of interconnected cells embedded in bone matrix, called osteocytes. There has been a growing interest in osteocytes in recent years with the realisation that these mechano-sensing cells orchestrate bone-resorbing and bone-forming processes involved in bone adapation and repair~\cite{franz-hall-witten,dallas-bonewald,dallas-prideaux-bonewald,marotti-2000,hughes-petit,sims-bonekey,rochefort-benhamou}. Osteocytes transduce mechanical stimuli, such as local deformations of the bone matrix, into biochemical signals transmitted to bone-resorbing cells (osteoclasts) and bone-forming cells (osteoblasts) through the bone surface. The ubiquitous role of osteocytes in the regulation of bone tissues is evidenced by several recent experimental studies. Osteocytes participate in regulating bone formation, in particular through secretion of sclerostin, an inhibitor of the $\wnt$ signalling pathway~\cite{dallas-bonewald,dallas-prideaux-bonewald,power-etal,atkins-etal-scl,sims-chia}. They are also known to produce $\rankl$ which initiates bone resorption through the promotion of osteoclastogenesis~\cite{dallas-prideaux-bonewald,nakashima-takayanagi-etal-2011,xiong-jilka-manolagas-etal-2011,obrien-nakashima-takayanagi-2013,sims-bonekey}. Osteocytes also help mineralise the soft matrix synthesised by the osteoblasts, and they regulate the degree of mineralisation of bone~\cite{atkins-findlay,atkins-etal-scl,barragan-adjemian-bonewald-etal,dallas-prideaux-bonewald}.

\begin{figure}[t] \centering\includegraphics[width=0.5\figurewidth]{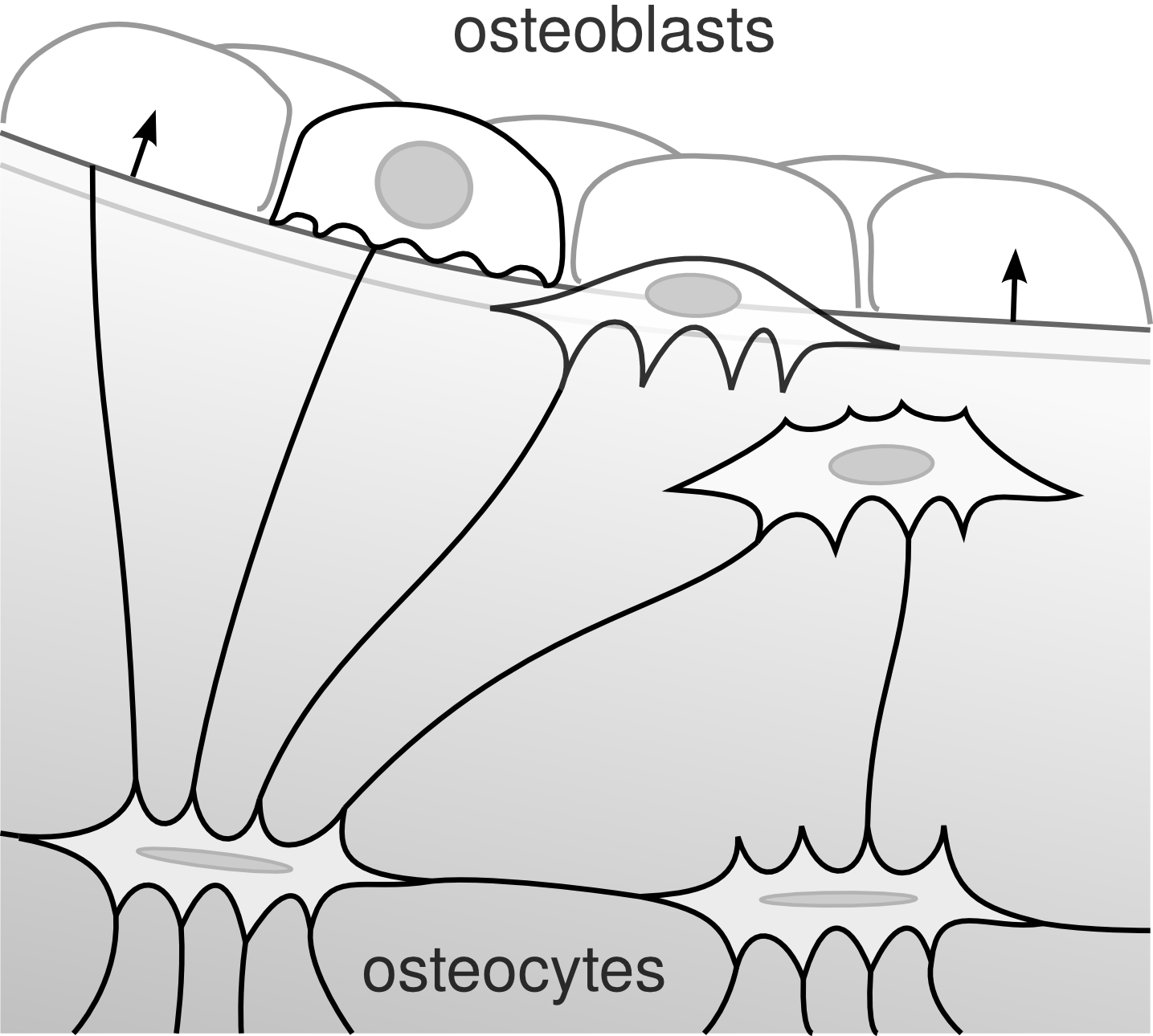}
    \caption{Osteocytes differentiate from osteoblasts that become trapped in bone matrix during bone deposition. They develop dendritic cell processes that connect to other osteocytes and to cells at the bone surface.}
    \label{fig:ocy-schematic}
\end{figure}

The network of osteocytes in bone matrix is generated during new bone deposition when some of the bone-forming osteoblasts become trapped and buried in the synthesised matrix (Figure~\ref{fig:ocy-schematic}). These cells gradually change their appearance and phenotype, becoming first osteoid-osteocytes, before terminally differentiating into osteocytes~\cite{palumbo-marotti-etal-1990a,palumbo-marotti-etal-1990b,nefussi-etal,franz-hall-witten,marotti-1996,marotti-2000}. During the osteoblast-to-osteocyte transition, the cells develop several dendritic processes connecting to the layer of matrix-synthesising osteoblasts above and to nearby osteocytes. 

Few mathematical models have modelled explicitly the generation of osteocytes from a population of osteoblasts. Polig and Jee~\cite{polig-jee} and Buenzli \etal~\cite{buenzli-etal-bmu-refilling} have explicitly included the varying depletion of osteoblasts due to osteocyte generation so as to retrieve a constant~\cite{polig-jee} or an experimentally-determined~\cite{buenzli-etal-bmu-refilling,hannah-etal} osteocyte density distribution in cortical basic multicellular units ($\bmu$)s. In Ref.~\cite{martin-bucklandwright}, Martin and Buckland-Wright model a similar depletion of osteoblasts in a bone-forming microsite undergoing trabecular remodelling. A depletion of ten osteoblasts over the microsite area $6,500\,\um^2$ is assumed to occur at discrete intervals, \ie, when the depth of mineralised matrix reaches 15, 30, and 45\,$\um$. This discrete depletion models Marotti's hypothesis that osteocytes prompt the burial of osteoblasts when they become sufficiently covered with bone matrix~\cite{marotti-1996,marotti-2000}. In purely temporal settings, Moroz, Wimpenny~\etal~\cite{moroz-etal-2006,wimpenny-moroz} assume osteocytes to be generated at constant density in the matrix and removed in proportion to the level of mechanical stress, Ascolani and Li\`o~\cite{ascolani-lio} assume osteocytes to be generated in proportion to the number of osteoblasts and removed at a constant rate for one day after an explicit microfracture, and Graham, Ayati\ etal~\cite{graham-ayati-etal} consider a single remodelling event during which osteocytes are generated in proportion to the number of osteoblasts with a logistic rate of growth, and removed artificially at the start of the simulation to initiate the remodelling event. Several other models have included the effect of local mechanical stimuli onto bone remodelling events~\cite{mullender-huiskes,vanOers-etal-2008,vanOers-etal-2011,baiotto-etal,adachi-etal-2010,kwon-etal,wang-vanoers-etal,andreaus-colloca-iacoviello,pivonka-buenzli-etal-specific-surface}. In the discrete models of van Oers \etal~\cite{vanOers-etal-2008,vanOers-etal-2011}, coupled formation and resorption results from changes in local mechanical stimuli provoked by small resorption cavities, which create local stress concentrations. Other models include the influence of local damage on the renewal rate of bone matrix~\cite{garciaaznar-rueberg-doblare,wang-vanoers-etal}. However, the exact role of osteocytes in these models of mechanics-driven bone regulation is unclear. None of these models accounts explicitly for a population of osteocytes and their generation.

This paper proposes a mathematical model of osteocyte generation that accounts for the spatially localised process of osteoblast burial at the moving deposition front. The consideration of osteocyte generation in this proper spatio-temporal setting enables to set on a microscopic basis what features of bone formation dynamics determine the density of osteocytes in bone and remain imprinted as such as a record~\cite{stein-werner}. This approach contrasts with the purely temporal models of osteocyte generation referred to above which rely on distinct effective assumptions on osteocyte generation. The advantage of the present approach is that the microscopic, cell level variables of osteoblast burial are explicitly displayed. The model elucidates the interplays between osteoblast density, matrix secretory rate, rate of burial, and curvature of the bone substrate in determining the density of osteocytes in the new bone matrix. The model and its results are applicable to any bone surface morphology. They apply to formation during cortical bone remodelling, trabecular bone remodelling, bone modelling, and bone growth~\cite{martin-burr-sharkey}.

The results of the model are combined with experimental data to provide estimates of the rate of burial of osteoblasts during bone formation. To the author's knowledge, these estimates are novel. It is hoped that they will lead to new insights into the burial process of osteocytes in different individuals, skeletal sites, species, and bone disorders. These estimates also enable an investigation of the spatial range of the osteocytic signal controlling bone formation hypothesised by Marotti~\cite{marotti-1996,marotti-2000}.

\section{Mathematical model}
The model of osteocyte generation presented in this paper is based on cell population balance equations in which source and sink terms are defined as biochemical reaction rates involving continuous local cell densities. This approach has been used previously to develop a number of mathematical models of bone cell development in both temporal and spatiotemporal settings~\cite{lemaire-etal,pivonka-etal1,buenzli-etal-anabolic,pivonka-buenzli-etal-specific-surface,buenzli-etal-trabecularisation,buenzli-etal-moving-bmu,buenzli-etal-bmu-refilling}. This approach is suitable to describe osteocyte generation even considering the localised nature of the osteoblast burial process that occurs at the moving deposition front. Indeed, the material balance principle upon which this approach relies is valid beyond the continuum model and is understood in this paper in the sense of generalised functions (distributions) whenever required~\cite{evans-morriss,jones-distrib}. 

The consideration of local cell densities also enables direct comparisons with experimentally determined densities. New bone matrix is deposited by a layer of osteoblasts densely packed at the surface of the bone substrate~\cite{marotti-zallone-ledda,franz-hall-witten}. The population of the matrix synthesising cells will therefore be characterised here by the local osteoblast surface density $\rho_\ob$ (number of cells per unit surface), as measured in Refs~\cite{jones-ob,marotti-zallone-ledda,pazzaglia-etal-2014}. The population of osteocytes will be characterised by the local volumetric density of cells, $\ocy$ (number of cells per unit volume), as measured in Refs~\cite{hannah-etal,mader-mueller-etal,dong-peyrin-etal}. Note that it is the bone-formation-driven moving bone surface that generates a volumetric density of osteocytes from the surface density of osteoblasts (Figure~\ref{fig:ocy-schematic}). Generally, it is expected that the density of osteocytes is determined by a combination of:
\begin{itemize}
    \item The osteoblast surface density $\rho_\ob\ [\#/\mm^2]$;
    \item The rate of osteoblast burial $\dburial\ [\da^{-1}]$, \ie, the probability per unit time~\cite{vankampen} for a single osteoblast to become buried in the newly deposited matrix;
    \item The matrix secretory rate $\kform\ [\mm^3/\da]$, \ie, the volume of new matrix secreted per osteoblast per unit time (usually reported in $\um^3/\da$~\cite{jones-ob,marotti-zallone-ledda});
    \item The geometry of the bone substrate upon which new bone is deposited by the osteoblasts. Indeed substrate geometry (in particular local surface curvature) can strongly influence the evolution of the bone surface~\cite{rumpler-fratzl-etal,bidan-fratzl-dunlop-etal-plos1,bidan-fratzl-dunlop-etal-bis}.
\end{itemize}

In this paper, the osteoblast surface density $\rho_\ob$ is assumed to be known at each time and every point of the surface. This population could be either determined experimentally~\cite{jones-ob,marotti-zallone-ledda,pazzaglia-etal-2014} or taken from the output of mathematical models such as that of Buenzli~\etal~\cite{buenzli-etal-bmu-refilling}.

\subsection{Planar bone substrate}
For simplicity, it is assumed in this section that the bone substrate is planar and that osteoblasts are uniformly distributed over the bone surface, \ie, $\rho_\ob = \rho_\ob(t)$. Each cell is further assumed to synthesise new bone matrix (osteoid) at a uniform rate $\kform(t)$. (These assumptions are relaxed in the next section.) Clearly, the bone surface remains planar at all times and its evolution can be tracked by the new bone layer thickness $w(t)$ deposited over the initial substrate. During a small time increment $\Delta t$, the volume of new bone synthesised by a single osteoblast is $\kform(t)\Delta t$. Since a single osteoblast occupies an area $1/\rho_\ob$ on the surface, the new bone layer thickness is increased by $\Delta w = \kform\Delta t/\rho_\ob^{-1}$. In the limit $\Delta t\to 0$, this leads to:
\begin{align}\label{w}
    \fd{}{t} w(t) = \kform(t)\,\rho_\ob(t).
\end{align}

The average number of osteoblasts that become buried in bone matrix during $\Delta t$ is $\Delta N_\ob=\dburial\rho_\ob S\Delta t$, where $S$ is the bone surface area. Therefore the rate of depletion in osteoblast surface density induced by osteoblast burial is given by:
\begin{align}\label{ob1}
    \fd{\rho_\ob}{t}\Big)_\text{burial} \ =\ -\dburial(t)\,\rho_\ob(t).
\end{align}
(This depletion of osteoblasts due to the generation of osteocytes is assumed to be implicitly accounted for in the given function $\rho_\ob(t)$.)

All the osteoblasts buried in bone matrix are assumed to give rise to new osteocytes. Whilst a number of phenotypic changes describe the osteoblast-to-osteocyte transition~\cite{palumbo-marotti-etal-1990a,palumbo-marotti-etal-1990b,nefussi-etal,marotti-1996,marotti-2000}, here these changes are assumed to occur instantly, \ie, a cell counts as an `osteocyte' as soon as it is buried. The osteoblast-to-osteocyte transition is assumed to take place instantaneously at the moving deposition front positioned at $z=w(t)$, where the $z$ axis is normal to the bone surface. The local rate of generation of osteocytes is thus governed by:
\begin{align}\label{ocy1}
    \pd{}{t} \ocy(t, z) = \dburial(t)\, \rho_\ob(t)\ \deltaup\big(z-w(t)\big).
\end{align}
The Dirac delta factor in Eq.~\eqref{ocy1}~\cite{jones-distrib} accounts for the fact that burial only occurs at the bone interface $z = w (t)$. This factor is also responsible for turning the depletion rate in osteoblast \emph{surface density} into the production rate of \emph{volumetric density} of osteocytes.\footnote{In one spatial dimension, the Dirac distribution has inverse length units.} 

Only the creation of osteocytes is considered in Eq.~\eqref{ocy1}. In fact, this equation corresponds to the creation of osteocyte lacunae (the small pores in bone matrix in which osteocytes live), which remain imprinted in bone at the location of their creation~\cite{pazzaglia-etal-2010} even after the death of the osteocyte they contain. A depletion of the live osteocytes population due to cell death (apoptosis) will be considered in Section~\ref{sec:apoptosis}.

 Equations~\eqref{w} and~\eqref{ocy1} describe how osteocyte density depends on osteoblast density, burial rate, and matrix secretory rate. These equations can be integrated to provide a closed expression for osteocyte density. To deal with the singularity at the moving deposition front in Eq.~\eqref{ocy1}, one can introduce the final, $z$-dependent density of osteocytes
\begin{align}\label{ocyinf1}
    \ocy_\infty(z) \equiv\ocy(t\to\infty,z)
\end{align}
obtained once deposition has stopped or has moved far enough from the region of interest. Since no osteocyte is present initially for any location $z$ corresponding to newly deposited bone, $\ocy(t\!=\!0,z)=0$, and one obtains from Eq.~\eqref{ocy1}:
\begin{align}\label{ocyinf2}
    \ocy_\infty(z) = \!\int_0^\infty \!\!\!\!\!\!\der t\, \tpd{}{t}\ocy(t,z) = \!\int_0^\infty\!\!\!\!\!\!\der t\, \dburial(t)\,\rho_\ob(t)\, \deltaup\big(z\!-\!w(t)\big).
\end{align}
During bone formation, the width of the layer of new bone $w(t)$ is an increasing function of $t$. Hence, there is a unique time $t^\ast$ such that $z = w (t^\ast)$ and only the time $t = t^\ast$ contributes to the integral. Using $\deltaup\big(z-w(t)\big) = \deltaup(t-t^\ast)/|\td{}{t}w(t^\ast)|$~\cite{jones-distrib} and Eq.~\eqref{w} to substitute $\td{}{t}w(t^\ast)$, Eq.~\eqref{ocyinf2} gives:
\begin{align}\label{ocyinf-result1}
    \ocy_\infty\big(w(t^\ast)\big) = \frac{\dburial(t^\ast)}{\kform(t^\ast)}.
\end{align}
This result states that the density of osteocytes generated at the bone deposition front is simply the ratio of the burial rate and the matrix secretory rate. In particular, the density of osteocytes does not depend explicitly on the density of osteoblasts. Indeed, if there are few osteoblasts, there are few osteocytes generated per unit time, but there is also little matrix deposited. If there are many osteoblasts, there are many osteocytes generated but also a large amount of matrix deposited. These effects compensate themselves in determining the volumetric density of osteocytes. 

In fact, the result~\eqref{ocyinf-result1} can be checked from the following geometric argument. During the time increment $\Delta t$, a number $\Delta N_\ob=\dburial\rho_\ob S \Delta t$ of osteoblasts become osteocytes. These osteocytes are embedded in a total volume of new bone matrix $S\Delta w = \kform \rho_\ob S \Delta t$, and so the average volumetric density of osteocytes is given by $\frac{\Delta N_\ob}{S\Delta w}=\frac{\dburial}{\kform}$. The result~\eqref{ocyinf-result1} therefore provides a direct validation of the microscopic law of osteocyte generation, Eq.~\eqref{ocy1}.

\subsection{Nonplanar bone substrate}
Most bone matrix deposition during bone remodelling occurs on nonplanar substrates. In cortical bone, the resorption cavities opened up by the bone-resorbing cells during remodelling are roughly cylindrical, and thus, so is the bone substrate upon which osteoblasts subsequently deposit new bone. Trabecular bone deposition may occur on flatter surfaces, but complex curvatures are present at intersections of trabecular struts or plates~\cite{cowin-handbook}.

The deposition of bone matrix on nonplanar substrates defines an evolving bone surface $S(t)$ possessing space-dependent properties such as curvature. Since curvature may influence both the evolution of the osteoblast surface density and the matrix secretory rate ~\cite{buenzli-etal-bmu-refilling,marotti-zallone-ledda,qiu-parfitt-etal-2010,rumpler-fratzl-etal,bidan-fratzl-dunlop-etal-plos1,bidan-fratzl-dunlop-etal-bis,lim-etal}, it is important to account for potential spatial dependences in $\rho_\ob$ and $\kform$. As before, we assume known the surface density of osteoblasts $\rho_\ob(t,\b r_S)$, where $\b r_S$ is a point of the surface $S(t)$. 

The evolution of the bone surface due to matrix deposition is now governed by:
\begin{align}\label{v-normal}
    v(t, \b r_S) = \kform(t, \b r_S) \rho_\ob(t, \b r_S),
\end{align}%
\begin{figure}[t] \centering\includegraphics[width=0.5\figurewidth]{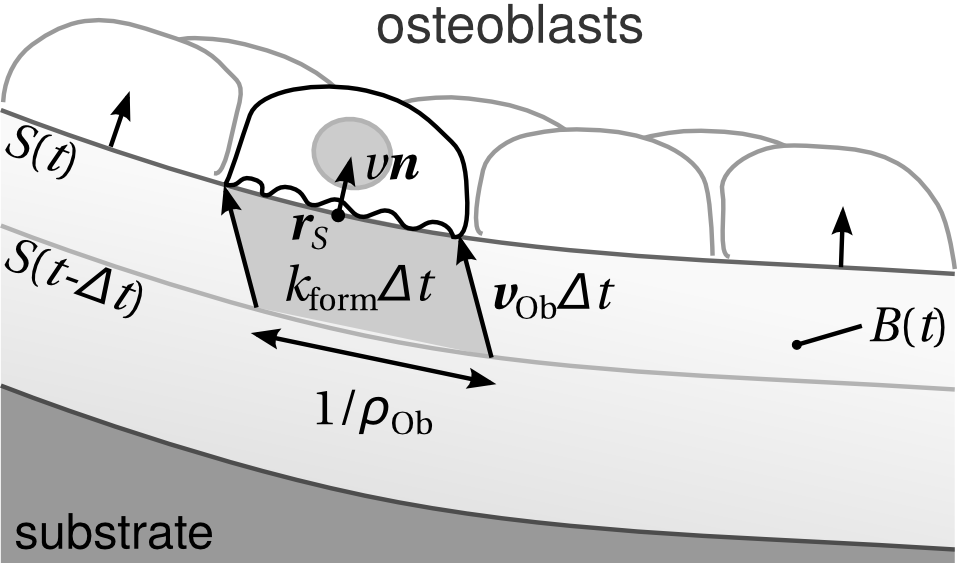}
    \caption{Evolution of the bone surface $S(t)$ due to new bone matrix deposition on a nonplanar substrate by osteoblasts. See text and Eq.~\eqref{v-normal} for more details.}
    \label{fig:bone-deposition}
\end{figure}%
where $v(t, \b r_S)$ is the normal velocity of the moving front $S(t)$ at $\b r_S$ (Figure~\ref{fig:bone-deposition}), referred to by biologists as the matrix apposition rate~\cite{martin-burr-sharkey,metz-martin-turner,buenzli-etal-bmu-refilling}. The expression~\eqref{v-normal} is found by equating the volume of matrix $\kform \Delta t$ secreted during $\Delta t$ by the osteoblast at $\b r_S$, with the volume $\rho_\ob^{-1} \b v_\ob\Delta t \cdot\b n$ of the region that this osteoblast `sweeps' on its path (shown as a darker gray shade in Figure~\ref{fig:bone-deposition}), where $\b n$ is the outward unit vector normal to $S(t)$ at $\b r_S$, and $\b v_\ob$ is the osteoblast's velocity vector. Note that the evolution of the bone surface is uniquely determined by the knowledge of the normal velocity only $v\equiv\b v_\ob\cdot\b n$ everywhere on the surface, see e.g.~\cite{sethian,osher-fedkiw}.

The natural generalisation of Equation~\eqref{ocy1} to nonplanar geometries is:
\begin{align}\label{ocy2}
    \pd{}{t}\ocy(t, \b r) = \dburial(t, \b r)\,\rho_\ob(t,\b r)\,\deltaup_{S(t)}(\b r),
\end{align}
where $\b r$ is a point in 3D space, and $\deltaup_{S(t)}(\b r)$ is the ``Dirac wall'' or surface Dirac distribution on $S(t)$, \ie, formally infinite anywhere on $S(t)$ and zero everywhere else, such that for any test function $\varphi$:
\begin{align}\label{dirac-wall-def}
    \int\der^3 r\ \varphi(\b r)\,\deltaup_{S(t)}(\b r) = \int_{S(t)}\!\!\!\der\sigma(\b r_S)\,\varphi(\b r_S),
\end{align}
where $\int_{S(t)}\!\der\sigma(\b r_S)$ is the line integral over $\b r_S\in S(t)$~\cite[Sec 8.4]{jones-distrib}. Since $\tpd{}{t}\ocy(t, \b r) = 0$ for $\b r\not\in S(t)$, $\ocy(t, \b r)$ is time-independent almost everywhere, \ie\ of the form:
\begin{align}\label{ocy-result2}
    \ocy(t, \b r) = \begin{cases}\ocy_\infty(\b r), & \b r\in B(t),
\\0, & \b r\not\in B(t) \end{cases} \equiv \ocy_\infty(\b r) \chi_{B(t)}(\b r),
\end{align}
where $\ocy_\infty(\b r) = \ocy(t\to\infty, \b r)$ is the final, space-dependent density of osteocytes, $B(t)$ is the region of space occupied by new bone at time $t$ (Figure~\ref{fig:bone-deposition}), and $\chi_{B(t)}$ is the indicator function of $B(t)$. Because the boundary $S(t)$ of $B(t)$ moves with normal velocity $v$, the quantity $\chi_{B(t+\Delta t)}(\b r)-\chi_{B(t)}(\b r)$ is nonzero only in a layer of thickness $|v(t,\b r)| \Delta t$ extending normally from $S(t)$. When $\Delta t\to 0$, this layer becomes infinitely thin
whilst the nonzero value diverges upon division by $v(t,\b r)\Delta t$, and one has:
\begin{align}\label{dirac-wall}
    \frac{\chi_{B(t +\Delta t )}(\b r) - \chi_{B(t )}(\b r)}{v(t,\b r)\Delta t} \to \deltaup_{S(t )}(\b r), \quad \Delta t \to 0
\end{align}
(see the Appendix). Differentiating Eq.~\eqref{ocy-result2} with respect to $t$, one thus obtains:
\begin{align}
        \pd{}{t}\ocy(t,\b r)=\ocy_\infty(\b r)\pd{}{t}\chi_{B(t)}(\b r)=\ocy_\infty(\b r)v(t,\b r)\deltaup_{S(t)}(\b r).
\end{align}
By comparison with Eq.~\eqref{ocy2} and substitution of $v$ using Eq.~\eqref{v-normal}, the density of osteocytes generated at the deposition front in nonplanar geometries is:
\begin{align}\label{ocyinf-result2}
    \ocy_\infty\big(\b r_S(t)\big) = \frac{\dburial\big(t, \b r_S(t)\big)}{\kform(t, \b r_S(t)\big)},
\end{align}
where $\b r_S(t)\in S(t)$ is a point of the bone surface at each time (\eg, a point following the trajectory of an osteoblast). The result~\eqref{ocyinf-result2} generalises Eq.~\eqref{ocyinf-result1} to arbitrarily curved surfaces. Neither the density of osteoblasts nor the curvature of $S(t)$ explicitly influence the density of osteocytes deposited at the moving front. Whilst deposition of matrix on a curved surface can greatly concentrate or dilute locally the matrix-synthesising cells due to the local contraction or expansion of the available surface~\cite{buenzli-etal-bmu-refilling}, it is sufficient that the ratio of burial rate and matrix secretory rate of individual osteoblasts is maintained to generate a uniform distribution of osteocytes. This is a valuable property in bone formation given the importance of osteocyte distribution in sensing the local mechanical state of bone matrix and the complex morphologies of bone microarchitectures.

The geometric argument sketched in the planar case can be repeated here. It shows that Eq.~\eqref{ocyinf-result2} is correct in the continuous limit, and/or at leading order in curvature. Indeed, the volume of new bone matrix formed by deposition of a layer of thickness $\Delta w$ normally to a small element of surface of area $\Delta\sigma$ is $\Delta V = \Delta\sigma\Delta w \big[1+\Order(\Delta\sigma,\Delta w)\big]$, where the subdominant orders arise in curved geometries. For instance for a cylinder of radius $R$ and axis $x$, one has $\Delta\sigma = R\Delta\theta\Delta x$ and $\Delta V = \frac{\Delta\theta}{2}\big((R+\Delta w)^2 - R^2\big)\Delta x = \Delta\sigma\Delta w \big(1+\frac{\Delta w}{2R}\big)$. With the number of osteocytes generated within this volume $\Delta N_\ob = \dburial \rho_\ob\Delta\sigma\Delta t$, and with $\Delta w = \kform \rho_\ob\Delta t$, the volumetric density is:
\begin{align}\label{ocyinf-result3}
    \frac{\Delta N_\ob}{\Delta V} = \frac{\dburial\rho_\ob\Delta \sigma\Delta t}{\Delta \sigma\Delta w\big[1+\Order(\Delta\sigma, \Delta w)\big]}=\frac{\dburial}{\kform}+\Order(\Delta\sigma,\Delta w).
\end{align}
The result~\eqref{ocyinf-result2} therefore validates the microscopic law of osteocyte generation in curved geometries postulated in Eq.~\eqref{ocy2}. Whilst the microscopic law~\eqref{ocy2} is not strictly necessary to deduce what dynamic processes of bone formation determine the density of osteocytes~\eqref{ocyinf-result2}, knowing how to formulate osteocyte generation in a microscopic spatio-temporal framework enables the straightforward inclusion of further processes known to occur at this level. This is particularly advantageous for including coupling with other processes governed by the material conservation law. This will be illustrated in Section~\ref{sec:apoptosis} by including osteocyte apoptosis (programmed cell death) in Eq.~\eqref{ocy2}.

\paragraph{Implicit dependences of osteocyte density}
The fact that neither osteoblast density nor curvature appear in Eq.~\eqref{ocyinf-result2} may seem surprising. A number of experimental studies suggest that osteocyte density correlates with osteoblast density~\cite{vashishth-fyhrie-etal,qiu-parfitt-etal-abstract,zarrinkalam-atkins-etal}. However, this is not always the case~\cite{pazzaglia-etal-2012}. Likewise, osteocyte density in rapidly-laid woven bone can be similar to osteocyte density in slower-laid lamellar bone, or higher, depending on skeletal site and developmental history~\cite{hernandez-majeska-schaffler}. Generally speaking, osteocyte density may correlate with several variables during bone formation, including hormones, skeletal site, distance to bone surface, mechanical environment, and species~\cite{qiu-parfitt-etal-2002,qiu-parfitt-etal-2002b,metz-martin-turner,zarrinkalam-atkins-etal,hernandez-majeska-schaffler,sharma-etal,hannah-etal,carter-thomas-clement-etal,pazzaglia-etal-2012,dong-peyrin-etal,mullender-huiskes-etal-1996,mader-mueller-etal,stein-werner}.

It has to be emphasised that both the burial rate $\dburial$ and the matrix secretory rate $\kform$ may implicitly depend upon osteoblast density, curvature, and other dynamic variables regulating the bone formation process, including site-specific signalling molecules produced by underlying osteocytes in response to mechanical stimuli:
\begin{align}
    \dburial&(\rho_\ob, \ocy, \text{curvature, mech.}, \ldots) \notag
    \\\kform&(\rho_\ob, \ocy, \text{curvature, mech.}, \ldots) \notag
\end{align}
It is these implicit dependences that are responsible for the observed variations in osteocyte density in the aforementioned studies. Provided one knows the burial rate $\dburial$ and the matrix secretory rate $\kform$, osteocyte density is always determined by taking their ratio, Eq.~\eqref{ocyinf-result2}.

For example, crowding of osteoblasts such as due to bone surface shrinkage in cortical $\bmu$s~\cite{buenzli-etal-bmu-refilling}, and convex bone substrate geometries are likely to increase the probability of osteoblasts overlapping each other and burying their peers~\cite{franz-hall-witten,marotti-2000,nefussi-etal}, generating a dependence of osteocyte density upon osteoblast density through $\dburial$. Note, however, that additional concurrent mechanisms may strongly regulate the surface density of osteoblasts~\cite{marotti-zallone-ledda,pazzaglia-etal-2012}, such as anoikis (cell death induced by detachment from the bone surface), a process believed to explain the large difference between the total number of osteoblasts generated in a remodelling event and the residual number of osteocytes~\cite{martin-burr-sharkey,parfitt-1994,pazzaglia-etal-2014}.

It is clear however that the probability for an osteoblast to become buried must depend on matrix secretory rate, since no osteoblast can be buried if $\kform=0$:
\begin{align}
    \dburial(\kform\!=\!0) = 0.
\end{align}
Caution must be exerciced when taking a simple linear dependence between $\dburial$ and $\kform$ in a mathematical model such as Eq.~\eqref{ocy2}, as this amounts to setting the osteocyte density generated at the deposition front to the proportionality factor.

\section{Estimation of burial rate and of the range of osteocytic control of burials}
Little is known about the detailed mechanism by which osteoblasts become buried in bone matrix during matrix deposition~\cite{franz-hall-witten}. To the author's knowledge, the fact that the density of osteocytes is determined exactly by the ratio of osteoblast burial rate and matrix secretory rate has not been recognised and fully appreciated in the literature to date. In this section, some consequences of using Eq.~\eqref{ocyinf-result2} in conjunction with experimental data are therefore explored. They lead to new insights into the osteoblast burial process, such as (i) estimations of the rate of osteoblast burial during bone refilling in human cortical basic multicellular units ($\bmu$s) and at rabbit endosteal plates, and (ii) a determination of how burial rate may be controlled by the distribution of underlying osteocytes~\cite{marotti-1996,marotti-2000}.

\subsection{Osteoblast burial rate}\label{sec:dburial}
\paragraph{Burial rate in human intracortical bone formation} The result~\eqref{ocyinf-result2} enables the determination of the rate of burial of osteoblasts $\dburial$ from experimental measurements of osteocyte density and matrix secretory rates. 

\begin{figure}[t]    \includegraphics[width=\figurewidth]{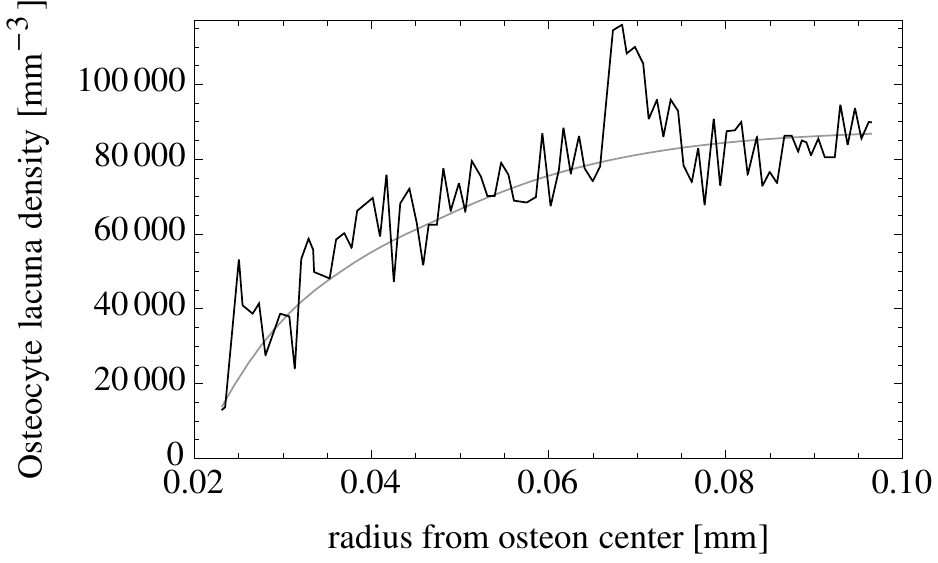}
    \caption{Radial dependence of the density of osteocytes $\ocy_\infty(R)$ in an osteon. Black: data from Ref.~\cite{hannah-etal}. Gray: smoothed interpolating curve.}
    \label{fig:ocy-vs-R}    
\end{figure}

In cortical bone, the renewal of bone matrix is perfomed by self-consistent groups of bone-resorbing and bone-forming cells called basic multicellular units ($\bmu$s)~\cite{sims-martin-2014}. This renewal process generates cylindrical structural elements of bone tissue called osteons~\cite{martin-burr-sharkey}. The distribution of osteocytes within an osteon is thus the record of a consistent and organised bone formation process. In a recent study, the spatial distribution of osteocyte lacunae in human osteons was investigated by imaging a femoral cortical bone sample with synchrotron-radiation micro-CT~\cite{hannah-etal}. These authors determined the radial dependence of osteocyte density within osteons,
$\ocy_\infty(R)$ (see Figure~\ref{fig:ocy-vs-R}).

\begin{figure}[t]
 \includegraphics[width=\figurewidth]{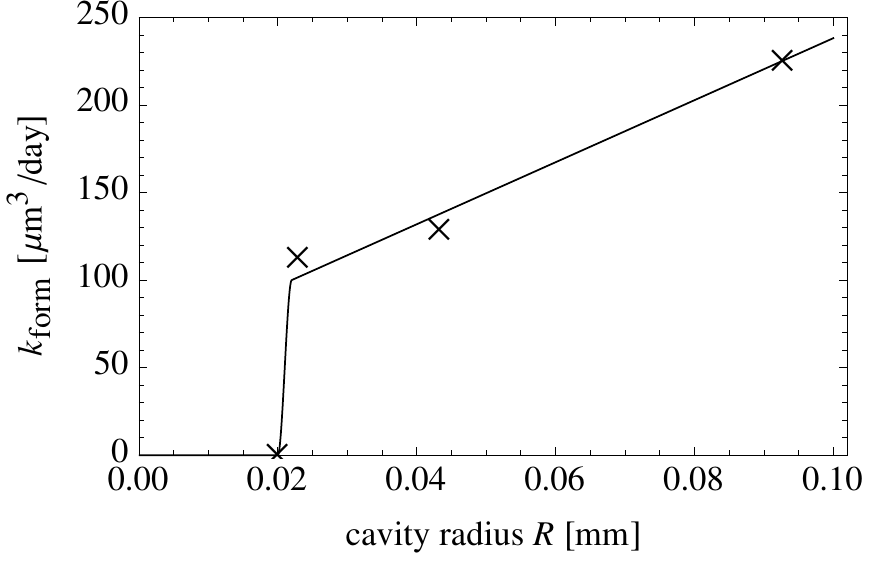}
    \caption{Matrix secretory rate $\kform$ as a function of resorption cavity radius. Crosses: data from~\cite{marotti-zallone-ledda}. Line: extrapolation (see~\cite{buenzli-etal-bmu-refilling} for more details).}
    \label{fig:kform-vs-R}    
\end{figure}
\begin{figure}[!t]
    \includegraphics[width=\figurewidth]{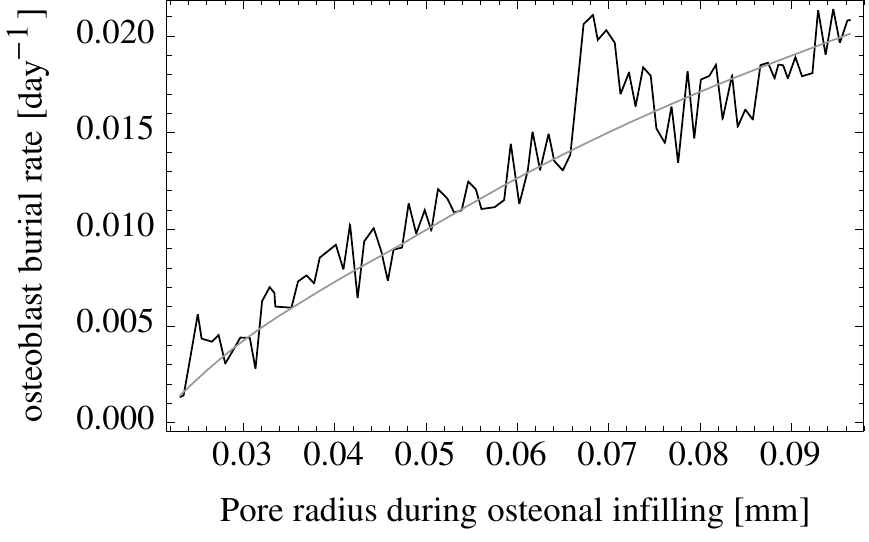}
    \caption{Burial rate of osteoblasts $\dburial(R)$ as a function of cavity radius. Black/gray: based on data from black/gray curve in Figure~\ref{fig:ocy-vs-R}.}
    \label{fig:dburial-vs-R}    
\end{figure}

The rate of matrix synthesis by individual osteoblasts $\kform$ can be experimentally determined from Eq.~\eqref{v-normal} by measuring both the matrix apposition rate $v$ (\eg, via tetracycline double labelling~\cite{metz-martin-turner}) and the surface density of osteoblasts~\cite{jones-ob,marotti-zallone-ledda,pazzaglia-etal-2014,buenzli-etal-bmu-refilling}. However, different experimental techniques are required to measure these quantities and few studies exist in which they are reported together. In Figure~\ref{fig:kform-vs-R}, experimental measurements from Ref.~\cite{marotti-zallone-ledda} of $\kform$ at various radii $R$ of dog cortical $\bmu$s are represented, correponding to different stages of osteonal infilling. These data are scaled to correspond to human cortical $\bmu$s and extrapolated into a continuous $R$-dependent function $\kform(R)$ following Ref.~\cite{buenzli-etal-bmu-refilling}, in which this function was used in a mathematical model of $\bmu$ validated against several independent measurements.

The rate of burial of osteoblasts $\dburial$ as a function of the radius of the closing $\bmu$ cavity can be estimated by combining these two types of data, $\ocy_\infty(R)$ and $\kform(R)$, into:
\begin{align}\label{dburial}
    \dburial(R) = \ocy_\infty(R)\ \kform(R).
\end{align}

By combining the data of Figures~\ref{fig:ocy-vs-R} and~\ref{fig:kform-vs-R} in Eq.~\eqref{dburial}, the average rate of burial of osteoblasts is seen to decrease as the cavity radius decreases, \ie\ as infilling of the resorption cavity proceeds in the $\bmu$ (see Figure~\ref{fig:dburial-vs-R}). This observation does not tell explicitly what biological processes may drive this progressive decrease in burial rate, but it shows that differences in burial rate can be large and can dominate differences in matrix secretory rate in determining osteocyte density.

\paragraph{Burial rate in rabbit endosteal bone formation}
The burial rate $\dburial$ introduced in the model is an overall measure of the probability per unit time for \emph{any} osteoblast to become buried. Pazzaglia \etal~\cite{pazzaglia-etal-2014} have estimated that approximately one in $67$ osteoblasts undergo an osteoblast-to-osteocyte transition over the course of deposition of a matrix layer the width of an osteocyte lacuna (\ie, about $5\,\um$ thick~\cite{dong-peyrin-etal}). This fraction of osteoblasts to become buried is implicitly included in the definition of $\dburial$. 

To estimate this fraction, Pazzaglia \etal\ compared in a rabbit model the surface density of osteoblasts and the surface density of open osteocyte lacunae visible on the bone surface after removal of the osteoblasts. These experimental measurements enables the estimation of $\dburial$ in this system from Eq.~\eqref{ocyinf-result2} as follows.

Let $w_\ocy$ be the width of an osteocyte lacuna in the direction of bone surface propagation, and let $\fracocyob$ be the ratio $\fracocyob\equiv\rho_\ocy/\rho_\ob\approx 1/67$, where $\rho_\ocy$ is the surface density of open osteocyte lacunae seen on the bone surface as measured by Pazzaglia \etal~\cite{pazzaglia-etal-2014}. One has $\ocy_\infty = \rho_\ocy/w_\ocy = \fracocyob\rho_\ob/w_\ocy$, and from Eq.~\eqref{v-normal}, $\kform\approx \frac{w_\ocy}{\rho_\ob T_\ocy}$, where $T_\ocy$ is the time required to deposit a layer of thickness $w_\ocy$. Substitution of these quantities into Eq.~\eqref{ocyinf-result2} leads to:
\begin{align}\label{dburial-fracocyob}
    \dburial = \ocy_\infty\, \kform = \fracocyob\frac{1}{T_\ocy}.
\end{align}
The factor $1/T_\ocy$ corresponds to the probability of burial per unit time of an osteoblast destined to become buried. Indeed, by definition the population of such osteoblasts has probability one to become buried within the layer of thickness $w_\ocy$ deposited during $T_\ocy$. Note that $1/T_\ocy=v/w_\ocy$.

Considering that the apposition rate $v$ in rabbit femur varies from about 2\,$\um/\da$ to 7\,$\um/\da$~\cite{cacchioli-etal-MAR-rabbit}, it takes $T_\ocy\approx 0.7$--$2.5\,\days$ to deposit a layer $w_\ocy\approx 5\,\um$ thick~\cite{dong-peyrin-etal}.\footnote{This estimate of $w_\ocy$ is made under the assumption that osteocyte lacuna dimensions are similar in rabbits than in humans.} During this period, a fraction $\fracocyob\approx 1/67$ of the osteoblasts will become buried. The burial rate $\dburial$~\eqref{dburial-fracocyob} during bone formation at the endosteal surface in rabbits therefore ranges from $0.006$--$0.02/\da$, which is in the same order as the values $\dburial$ in Figure~\ref{fig:dburial-vs-R} estimated in human cortical femurs.

\subsection{Osteocytic control of burial rate}
Marotti hypothesised that bone formation and burial of osteoblasts occur under tight control from the osteocytes in the bone matrix~\cite{marotti-1996,marotti-2000}. Osteocytes that find themselves buried deeper and deeper during matrix deposition may signal some osteoblasts at the surface to reduce their synthesising activity. These inhibited osteoblasts would then become buried by their peers~\cite{franz-hall-witten,nefussi-etal,marotti-2000}, and/or participate in their own burial through the development of a dorsal secretory territory~\cite{franz-hall-witten,pazzaglia-etal-2014}. Sclerostin is proposed as an important signalling molecule between osteocytes and osteoblasts. This signalling molecule secreted by osteocytes inhibits the $\wnt$ signalling pathway, and so osteoblast generation and activity~\cite{franz-hall-witten,power-etal,atkins-etal-scl,sims-chia}. Metz et al.~\cite{metz-martin-turner} have found that wall thickness in osteons is negatively correlated with osteocyte density, an indication that osteocyte-produced signals inhibit formation~\cite{vashishth-fyhrie-etal,atkins-findlay,zarrinkalam-atkins-etal,qiu-parfitt-etal-2002}. In the discrete computational models of Mullender and Huiskes and van Oers~\etal~\cite{mullender-huiskes,vanOers-etal-2008,vanOers-etal-2011}, bone microsites are implicitly assumed to contain osteocytes transducing the mechanical deformation of bone matrix into a mechanical stimulus such as sclerostin. The concentration of sclerostin is assumed to decrease exponentially from where it is produced with a characteristic decay distance of $100\,\um$, but the sclerostin concentration field can be increased by the multiplicity of osteocytes and their mechanical stimulation. Whilst these assumptions may be qualitatively reasonable~\cite{andreaus-colloca-iacoviello}, they are not based on experimental determinations of the reach of osteocyte signals onto bone cells. 
\begin{figure*}[t!] \includegraphics[width=\textwidth]{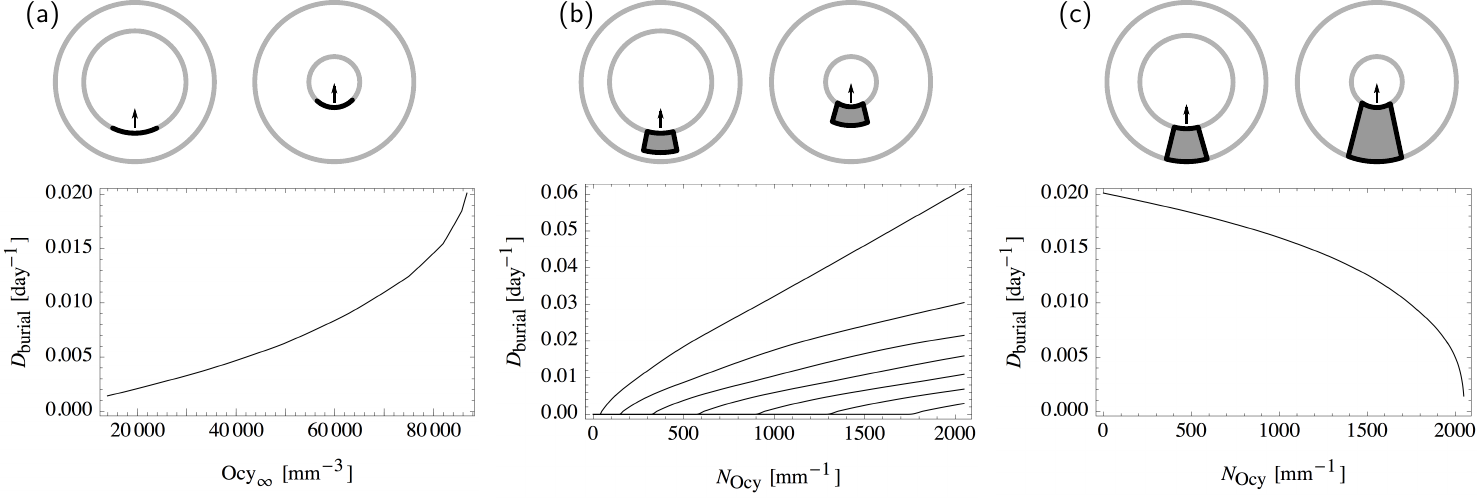}
    \caption{Correlations between burial rate and number of osteocytes in three distinct influence zones $\Omega(R)$ (seen as shaded areas of the $\bmu$ cross-section in the top row). The different curves in (b) correspond to influence zones of different fixed thickness $\delta R$. Cases (a) and (b) lead to a positive correlation between $\dburial$ and the number of osteocytes. Only (c) leads to a negative correlation consistent with Marotti's hypothesis.}
    \label{fig:dburial-vs-Nocy}
\end{figure*}

In this section, possible zones of influence of osteocytes onto osteoblast burial are examined. According to Marotti's hypothesis, the number of underlying osteocytes that help control the burial of new osteoblasts should correlate negatively with burial rate. Indeed, the more the number of underlying osteocytes, the less are needed, and therefore the smaller the burial rate. Such correlations can be investigated from the results of Section~\ref{sec:dburial} on $\dburial$ and the knowledge of osteocyte density.

The question investigated here is whether the overall decrease in burial rate $\dburial$ seen in Figure~\ref{fig:dburial-vs-R} as refilling proceeds may be attributed to a local signal proportional to the local density of osteocytes, or an integrated signal proportional to the total number of osteocytes found under the bone surface. In the following, a cylindrical cortical $\bmu$ geometry is assumed with negligible variation in the longitudinal axis of the $\bmu$. Let $\Omega(R)$ denote the influence zone of osteocytes in the $\bmu$ cross-section, \ie, the zone in which osteocytes help control osteoblast burial, where $R$ is the $\bmu$ resorption cavity radius (see Figure~\ref{fig:dburial-vs-Nocy}, top row). The number of osteocytes in this influence zone per unit longitudinal length is
\begin{align}
    N_\ocy(R) = \int_{\Omega(R)} \der A\ \ocy_\infty.
\end{align}
Below, the correlations between $\dburial$ and $N_\ocy$ are determined in three distinct influence zones $\Omega(R)$: (i) an infinitely-thin layer close to the bone surface, (ii) a layer of fixed width below the bone surface, and (iii) the full wall thickness of newly formed bone (Figure~\ref{fig:dburial-vs-Nocy}). 

\paragraph{Signal from the nearest osteocytes only} No osteocytes are formally present in an infinitely-thin layer, but one can investigate whether the rate of osteoblast burial at the deposition front (radius $R$) may be determined primarily by the density of the nearest existing osteocytes in the matrix, \ie, by $\ocy_\infty(R)$. Inverting $\ocy_\infty(R)$ as $R(\ocy_\infty)$
(smoothed curve in Figure~\ref{fig:ocy-vs-R}) and substituting into $\kform$ in Eq.~\eqref{dburial} exhibits the effective dependence of the burial rate upon the nearest osteocyte density:
\begin{align}
    \dburial(\ocy_\infty) = \ocy_\infty\ \kform(\ocy_\infty)
\end{align}
One sees from Figure~\ref{fig:dburial-vs-Nocy}a that burial rate increases when the density of nearby osteocytes increases. This observation is in conflict with Marotti's hypothesis. An osteocytic signal prompting osteoblast burial is therefore unlikely to arise solely from the closest osteocytes.

\paragraph{Local osteocytic signal} 
The rate of osteoblast burial may be determined by the superposition of signals emitted by a larger group of osteocytes beneath the bone surface, such as osteocytes within a layer of constant thickness $\delta R$:
\begin{align}
    N_\ocy(R) = 2\pi \int_{R-\delta R}^{R}\der r\ r\ \ocy_\infty(r).
\end{align}
Inverting $N_\ocy(R)$ as $R(N_\ocy)$ and substituting into $\ocy_\infty$ and $\kform$ in Eq.~\eqref{dburial} gives:
\begin{align}
        \dburial(N_\ocy) = \ocy_\infty(N_\ocy) \kform(N_\ocy).
\end{align}
One sees from Figure~\ref{fig:dburial-vs-Nocy}b that burial rate increases when the total number of osteocytes present in the influence zone increases, for any fixed thickness $\delta R$. This observation is again in conflict with Marotti's hypothesis.

\paragraph{Integrated osteocytic signal} The rate of osteoblast burial may be determined by the superposition of signals emitted by all the osteocytes found beneath the bone surface:
\begin{align}
    N_\ocy(R) = 2\pi \int_R^{R_c}\der r\ r\ \ocy_\infty(r),
\end{align}
where $R_c$ is the radius of the osteon boundary (cement line radius). By inversion and substitution into $\ocy_\infty$ as before, 
one sees from Figure~\ref{fig:dburial-vs-Nocy}c that the rate of burial decreases when the total number of osteocytes present increases. This observation is consistent with the hypothesis that low numbers of osteocytes will prompt the generation of new osteocytes, a likely regulation mechanism of osteocyte generation. These results suggest that osteoblasts may be influenced by integrated signals emanating from a large collection of underlying osteocytes. This emphasises the importance of understanding the role of the interconnectivity of the osteocyte network for signal transduction~\cite{adachi-etal-2009}. These observations give some support to the value of decay distance of $100\,\um$ for sclerostin chosen in the model by van Oers~\etal~\cite{vanOers-etal-2011}, as this distance is commensurate with an osteon's radius.

\section{Osteocyte apoptosis}\label{sec:apoptosis}
One of the benefits of the microscopic evolution law for osteocyte density~\eqref{ocy2} over the closed-form expression~\eqref{ocyinf-result2} is that Eq.~\eqref{ocy2} expresses the general conservation law of cell balance.  Hence further biochemical processes influencing osteocyte density can be accounted for in a consistent and modular fashion. To illustrate this, a term accounting for osteocyte apoptosis (cell death) is included in this section, and its consequence for osteocyte density is investigated.

Osteocytes are amongst the longest living cells in the body~\cite{franz-hall-witten,dallas-prideaux-bonewald}. Yet with age, the number of live osteocytes declines, which leaves a number of osteocyte lacunae empty. Osteocyte apoptosis is known to be increased by estrogen deficiency at menopause in women, and by age-related changes in testosterone in men~\cite{jilka-etal,dallas-prideaux-bonewald,sharma-etal}. As such, osteocyte apoptosis is an important process to account for in models of age-related bone loss~\cite{pivonka-buenzli-etal-specific-surface,buenzli-etal-trabecularisation}.

Apoptosis can be implemented in the model through the rate of osteocyte death $A(t, \b r)$, \ie, the probability per unit time for a single osteocyte at position $\b r$ to undergo cell death. This rate may generally be nonconstant and inhomogeneous. Its effect on the rate of change in osteocyte density is specified by conventional first order mass action kinetics, which modifies Eq.~\eqref{ocy2} into:
\begin{align}\label{ocy3}
    \pd{}{t}\ocy(t, \b r) = \dburial(t, \b r)\rho_\ob(t, \b r)\delta_{S(t)}(\b r) - A(t, \b r) \ocy(t, \b r).
\end{align}
Eq.~\eqref{ocy3} can be solved by use of the following Ansatz:
\begin{align}\label{ocy3-ansatz}
    \ocy(t, \b r) = \chi_{B(t)}(\b r) \ocy_\infty(\b r) \fracocylac(t, \b r),
\end{align}
where $\chi_{B(t)}$ is the indicator function of the region of new bone $B(t)$, $\ocy_\infty$ is the terminal osteocyte lacuna density, and $\fracocylac$ is the fractional occupancy of osteocyte lacunae, \ie, the fraction of lacunae occupied by live osteocytes. By differentiating with respect to $t$ and using Eqs~\eqref{v-normal}, \eqref{dirac-wall}, and \eqref{ocyinf-result2}, it can be shown that \eqref{ocy3-ansatz} is solution of Eq.~\eqref{ocy3} provided that $\fracocylac$ satisfies:
\begin{align}
    &\pd{}{t}\fracocylac(t, \b r) = - A(t, \b r) \fracocylac(t, \b r)\label{fracocylac-eq}
\\&\fracocylac\big(t, \b r_S(t)\big) = 1, \qquad \forall\b r_S(t)\in S(t), \forall t \label{fracocylac-bc}
\end{align}
The boundary condition~\eqref{fracocylac-bc} represents the assumption that each new lacuna generated at the moving deposition front $S(t)$ contains initially a live osteocyte. 

For planar geometries of the bone surface, the solution of Eq.~\eqref{fracocylac-eq} is
\begin{align}\label{fracocylac-sol}
    \fracocylac(t,z) = c(z) \exp{-\int_0^t\der\tau A(\tau)},
\end{align}
where the unknown function $c(z)$ is determined by the boundary condition~\eqref{fracocylac-bc} by evaluating Eq.~\eqref{fracocylac-sol} at $z=w(t)$. This leads to:
\begin{align}
    c\big(w(t)\big) = \e^{\int_0^t\der\tau A(t)}.
\end{align}
Using $t=w^{-1}(z)$ and Eq.~\eqref{ocyinf-result2}, the solution of Eq.~\eqref{ocy3} in planar geometries is therefore:
\begin{align}\label{ocy3-result-planar}
    \ocy(t, z) = \thetaup\big(w(t)-z\big)\frac{\dburial\big(w^{-1}(z)\big)}{\kform\big(w^{-1}(z)\big)} \exp\left\{-\int_{w^{-1}(z)}^{t}\hspace{-3.5ex}\!\der\tau\, A(\tau)\right\},
\end{align}
where $\thetaup$ is the Heaviside step function. If the apoptosis rate $A$ is constant, $\ocy(t,z)$ describes a wave progressing at speed $\fd{w}{t}(t)$.

For arbitrarily curved bone surfaces, the problem~\eqref{fracocylac-eq}--\eqref{fracocylac-bc} is solved in the same way. If $v>0$ at all times, then for every point $\b r\in B(t)$, there is a unique time $t^\ast=t^\ast(\b r)\leq t$ such that $\b r\in S(t^\ast)$ and the general solution of Eq.~\eqref{ocy3} is:
\begin{align}\label{ocy3-result}
    \ocy(t, \b r) = \chi_{B(t)}(\b r) \frac{\dburial\big(t^\ast(\b r), \b r\big)}{\kform\big(t^\ast(\b r),\b r\big)} \exp\left\{-\int_{t^\ast(\b r)}^t\hspace{-2.5ex}\der\tau\, A(\tau, \b r)\right\}.
\end{align}
The time $t^\ast(\b r)$ corresponds to the time of osteoblast burial at $\b r$. The explicit time dependences of osteocyte density $\ocy(t, \b r)$ in the right hand side of Eq.~\eqref{ocy3-result} describe two distinct dynamic processes: (i) osteocyte are found in an evolving domain $B(t)$; and (ii) osteocytes have a probability to undergo apoptosis counting from the time $t^\ast$ of their creation.

When the ratio $\dburial/\kform$ is maintained constant, the density of osteocyte lacunae generated is constant irrespective of the time evolution of the bone surface. In constrast, the density of live osteocytes at $\b r$ explicitly depends on the moment of their creation $t^\ast(\b r)$, which is determined by the dynamics of the moving deposition front. This dynamics is directly influenced by osteoblast density, and so by curvature~\cite{buenzli-etal-bmu-refilling}. The distribution of occupancy of osteocyte lacunae therefore records some details of the spatio-temporal evolution of the bone surface which are not recorded in the distribution of osteocyte lacunae.

\section{Conclusions}
This paper proposes a model of osteocyte generation to investigate the osteoblast-to-osteocyte transition in a spatio-temporal setting. The dynamic factors of the bone forming process that remain recorded in the density of osteocyte lacunae are determined and some consequences of this relationship are explored.

Mineralised bone matrix is extremely stable and can subsist for thousands of years. The analysis of morphological and material properties of bone offers a window into bone formation processes of current and extinct animals. Particularly osteocytes are a promising avenue for analysing bone disorders or for paleobiological studies due to (i) their primordial role as biosensor of local mechanical strains, and (ii) their participation in the orchestration of bone remodelling. For example, osteocyte lacunae have been shown to contain information on growth rates and muscle attachment sites of extinct species~\cite[and Refs cited therein]{stein-werner}.

Because bone formation proceeds linearly by the gradual deposition of new bone on existing bone surfaces, some features of the dynamic processes occurring at the moving deposition front become `imprinted' in the bone matrix. Whilst osteocytes do not outlive an organism, their lacunae remain as footprints of their burial. It is often believed that the density of osteocytes generated is directly dependent on the density of osteoblasts~\cite{qiu-parfitt-etal-abstract,ascolani-lio}. The results presented here show instead that only the rate of osteoblast burial (probability per unit time for an osteoblast to get buried) and the secretory rate of bone matrix (volume of matrix secreted per osteoblast per unit time) determine osteocyte density, irrespective of osteoblast density and substrate curvature.

To the author's knowledge, the model of osteocyte generation presented in this paper is novel on at least two levels: (i) by accounting for the moving front of bone deposition and of osteocyte generation, and (ii) by considering arbitrary substrate geometries. The simple and intuitive results obtained provide a validation of the assumptions of the model formulated at the cell level.

The model has enabled for the first time estimations of the rate of burial of osteoblasts in bone matrix (see Figure~\ref{fig:dburial-vs-R} and Eq.~\eqref{dburial-fracocyob}). These estimations are based on experimental determinations of osteocyte densities and matrix secretory rate per osteoblast. Future such estimations can provide insights into cell-level mechanisms of osteocyte generation that may be different in different individuals, skeletal sites, species, and bone disorders~\cite{vashishth-fyhrie-etal,qiu-parfitt-etal-abstract,zarrinkalam-atkins-etal,pazzaglia-etal-2012,hernandez-majeska-schaffler,qiu-parfitt-etal-2002,qiu-parfitt-etal-2002b,metz-martin-turner,sharma-etal,hannah-etal,carter-thomas-clement-etal,pazzaglia-etal-2012,dong-peyrin-etal,mullender-huiskes-etal-1996,mader-mueller-etal,stein-werner}. 

Furthermore, Marotti's hypothesis of osteocytic control of osteoblast burial~\cite{marotti-1996,marotti-2000} was investigated for its consistency with different possible zones of osteocytic influence. How osteocytes control bone formation remains poorly understood. In average, osteoid-osteocytes are connected with 5--6 different osteoblasts through more than 20 dendritic processes~\cite{kamioka-etal}, whilst osteocytes are connected to one another through more than 80 dendritic processes~\cite{sharma-etal}. Our analysis suggests that an osteocytic signal to osteoblasts must integrate a large number of osteocytes to be consistent with a negative correlation between burial rate and number of osteocytes in an influence zone, \ie, such an osteocytic signal is likely to be nonlocal. The highly interconnected network of osteocytes is compatible with this analysis.

New dynamic imaging techniques have recently been developed that enable live observations of osteocyte burial \emph{in vitro}~\cite{dallas-bonewald,sims-bonekey}. These techniques may be able to shed light on some poorly understood mechanisms of osteoblast burial at the cell level, which will be useful for future refinements of the mathematical model developed in this paper.

\subsection*{Acknowledgements}
I would like to thank John G.\ Clement, C.\ David L.\ Thomas, Gerald\ J.\ Atkins, and Natalie A. Sims for fruitful discussions. I gratefully acknowledge the Australian Research Council for being the recipient of a Discovery Early Career Researcher Fellowship (project number DE130101191).

\begin{appendices}
\section*{Appendix: Governing equation of the domain indicator function}
In this appendix, the governing equation of the indicator function of an evolving domain is derived, and a justification of Eq.~\eqref{dirac-wall} is provided:
\begin{align}
    \frac{\chi_{B(t +\Delta t )}(\b r) - \chi_{B(t )}(\b r)}{v(t,\b r)\Delta t} \to \deltaup_{S(t )}(\b r), \quad \Delta t \to 0\notag\tag{\ref{dirac-wall}},
\end{align}
where $\deltaup_{S(t)}$ is the ``Dirac wall'', or surface Dirac distribution defined by Eq.~\eqref{dirac-wall-def}. An intuitive derivation of the governing equation of $\chi_{B(t)}$ is based upon the observation that in the limit $\der t\to 0$:
\begin{align}
    \chi_{B(t+\der t)}(\b r) = 
    \begin{cases} 
            1 &\text{if $\b r - \b v \der t\in B(t)$,}
            \\0 &\text{otherwise} 
    \end{cases}
    =\chi_{B(t)}(\b r-\b v \der t),
\end{align}
where $\b v(t, \b r)$ is the velocity of the boundary $S(t)$ of $B(t)$. Hence, $\chi_{B(t+\der t)}(\b r) - \chi_{B(t)}(\b r) \sim -\b v\cdot\b\nabla \chi_{B(t)}(\b r) \der t$ as $\der t\to 0$, leading to:
\begin{align}\label{indicator-eq}
    &\pd{}{t}\chi_{B(t)} + \b v \cdot \b\nabla \chi_{B(t)}=0.
\end{align}
An alternative approach to deriving Eq.~\eqref{indicator-eq} is to consider a level set function $\phi(t, \b r)$ whose zero level contour $\phi(t, \b r)=0$ describes the set $S(t)$~\cite{sethian,osher-fedkiw}, and then to differentiate $\chi_B(t)(\b r) \equiv \thetaup\big(\phi(t, \b r)\big)$ with respect to~$t$. Equation~\eqref{indicator-eq} is the same as that satisfied by the level function $\phi(t, \b r)$, except that it must be interpreted in the sense of generalised functions since the gradient of the indicator function $\chi_{B(t)}$ is singular on $S(t)$. To investigate the nature of this singularity, let us integrate $\b v\cdot\b\nabla \chi_{B(t)}$ with a test function $\varphi(\b r)$ of compact support~\cite{jones-distrib}, and use the identity:
\begin{align}
    \varphi \b v\cdot\b\nabla\chi_{B(t)} = -\b\nabla\cdot(\chi_{B(t)}\varphi\b v)-\chi_{B(t)}\b\nabla\cdot (\varphi \b v)
\notag.
\end{align}
Because $\varphi$ has compact support, the first term in the right hand side gives zero after integration over $\mathbb{R}^3$ and use of the divergence theorem. Thus:
\begin{align}\label{indicator-dirac-wall1}
    \int\!\!\!\der^3r\ \varphi\b v\cdot\b\nabla\chi_{B(t)} = - \int_{B(t)}\hspace{-2ex}\der^3r\ \b\nabla\cdot(\varphi\b v) = - \int_{S(t)}\hspace{-2ex}\der\sigma \b n\cdot \b v \varphi,
\end{align}
where $\b n$ is the outward unit normal vector to $S(t)=\p B(t)$ and the divergence theorem is used for the second equality. Since the normal velocity of $S(t)$ is $v=\b n\cdot \b v$, Eq.~\eqref{indicator-dirac-wall1} shows that:
\begin{align}\label{indicator-dirac-wall2}
    -\b v \cdot \b \nabla \chi_{B(t)} = v \deltaup_{S(t)}
\end{align}
(in the sense of distributions). Eqs~\eqref{indicator-eq} and~\eqref{indicator-dirac-wall2} justify Eq.~\eqref{dirac-wall}. The reader is referred to Refs~\cite[Sec 1.5]{osher-fedkiw}, \cite[Sec 8.4]{jones-distrib}, and \cite[Sec 4.1]{lange} for related developments.
\end{appendices}


\begin{thebibliography}{99}
\setlength{\itemsep}{-1ex}
\small
\bibitem{martin-burr-sharkey}
Martin RB, Burr DB and Sharkey NA (1998) \textit{Skeletal tissue mechanics} (New York: Springer)

\bibitem{cowin-handbook} Cowin SC (Ed) (2001) \emph{Bone mechanics handbook}, $2^\text{nd}$Ed. (CRC Press)

\bibitem{franz-hall-witten}
Franz-Odendaal TA, Hall BK, Witten PE (2006) Buried alive: how osteoblasts become osteocytes, \emph{Develop Dyn} 235:176

\bibitem{dallas-bonewald} Dallas SL, Bonewald LF (2010) Dynamics of the transition from osteoblast to osteocyte, \emph{Ann N Y Acad Sci}1192:437--443

\bibitem{dallas-prideaux-bonewald} Dallas SL, Prideaux M, Bonewald LF (2013) The osteocyte: an endocrine cell $\ldots$ and more, \emph{Endocr Rev} 34:658--690

\bibitem{marotti-2000}
Marotti G (2000) The osteocyte as a wiring transmission system, \emph{J Muskuloskel Neuron Interact} 1:133

\bibitem{hughes-petit} Hughes JM, Petit MA (2010) Biological underpinnings of Frost's mechanostat thresholds: The important role of osteocytes, \emph{J Musculoskelet Neuronal Interact} 10:128--135

\bibitem{sims-bonekey} Sims NA (2013) New insights into osteocyte and osteoblast biology: support of osteoclast formation, PTH action and the role of Wnt16 (ASBMR 2013), \emph{IBMS BoneKEy} 10:article 467

\bibitem{rochefort-benhamou} Rochefort GY, Benhamou C-L (2013) Osteocytes are not only mechanoreceptive cells, \emph{Int J Numer Meth Biomed Eng} 29:1082--1088

\bibitem{power-etal} Power, J, Poole KES, van Bezooijen R, Doube M, Caballero-Al\'ias A, Lowik C, Papapoulos S, Reeve J, Loveridge N (2010) Sclerostin and the regulation of bone formation: Effects in hip osteoarthritis and femoral neck fracture, \emph{J Bone Miner Res} 25:1867--1876

\bibitem{atkins-etal-scl} Atkins GJ, Rowe PS, Lim HP, Welldon KJ, Ormsby R, Wijenayaka AR, Zelenchuk L, Evdokiou A, Findlay DM (2011) Sclerostin is a locally acting regulator of late-osteoblast/preosteocyte differentiation and regulates mineralization through a MEPE-ASARM-dependent mechanism, \emph{J Bone Miner ResJ} 26:1425--1436

\bibitem{sims-chia} Sims NA, Chia LY (2012) Regulation of sclerostin expression by paracrine and endocrine factors, \emph{Clin Rev Bone Miner Metab} 10:98--107

\bibitem{nakashima-takayanagi-etal-2011} Nakashima T, Hayashi M, Fukunaga T et al.~(2011) Evidence for osteocyte regulation of bone homeostasis through RANKL expression, \emph{Nat Med} 17:1231--34

\bibitem{xiong-jilka-manolagas-etal-2011} Xiong J, Onal M, Jilka RL, Weinstein RS, Manolagas SC, O'Brien ChA (2011) Matrix-embedded cells control osteoclast formation, \emph{Nat Med} 17:1235--41

\bibitem{obrien-nakashima-takayanagi-2013} O'Brien ChA, Nakashima T, Takayanagi H (2013) Osteocyte control of osteoclastogenesis, \emph{Bone} 54:258--263

\bibitem{atkins-findlay}
Atkins GJ and Findlay DM (2012) Osteocyte regulation of bone mineral: a little give and take, \emph{Osteoporos Int} 23:2067--2079

\bibitem{barragan-adjemian-bonewald-etal} Barragan-Adjemian C, Nicolella D, Dusevich V, Dallas MR, Eick JD, Bonewald LF (2006) Mechanism by which MLO-A5 late osteoblasts/early osteocytes mineralize in culture: Similarities with mineralization of lamellar bone, \emph{Calcif Tissue Int} 79:340--353

\bibitem{palumbo-marotti-etal-1990a} Palumbo C, Palazzini S, Marotti G (1990) Morphological study of intercellular junctions during osteocyte differentiation, \emph{Bone} 11:401--406

\bibitem{palumbo-marotti-etal-1990b} Palumbo C, Palazzini S, Zaffe D, Marotti G (1990) Osteocyte differentiation in the tibia of newborn rabbit: An ultrastructural study of the formation of cytoplasmic processes, \emph{Acta Anat} 137:350--358

\bibitem{nefussi-etal} Nefussi JR, Sautier JM, Nicolas V, Forest N (1991) How osteoblasts become osteocytes: A decreasing matrix forming process, \emph{J Biol Buccale} 19:75--82

\bibitem{marotti-1996} Marotti G (1996) The structure of bone tissues and the cellular control of their deposition, \emph{Ital J Anat Embryol} 101:25--79

\bibitem{polig-jee} 
Polig E and Jee W S S (1990) A model of osteon closure in cortical bone. \textit{Calcif. Tissue Int.} \textbf{47}:261--269

\bibitem{buenzli-etal-bmu-refilling}
Buenzli PR, Pivonka P, Smith DW (2014), Bone refilling in cortical basic multicellular units: Insights into tetracycline double labelling from a computational model, \emph{Biomech Model Mechanobiol} 13:185--203

\bibitem{hannah-etal}
Hannah KM, Thomas CDL, Clement JG, De~Carlo F, Peele AG (2010) Bimodal distribution of osteocyte lacunar size in the human femoral cortex as revealed by micro-{CT}, \emph{Bone} 47:866

\bibitem{martin-bucklandwright} Martin MJ and Buckland-Wright JC (2005) A novel mathematical model identifies potential factors regulating bone apposition, \emph{Calcif Tissue Int} 77:250--260

\bibitem{moroz-etal-2006} Moroz A, Crane MC, Smith G, and Wimpenny DI (2006) Phenomenological model of bone remodeling cycle containing osteocyte regulation loop, \emph{BioSystems} 84:183--190

\bibitem{wimpenny-moroz} Wimpenny DI, Moroz A (2007) On allosteric control model of bone turnover cycle containing osteocyte regulation loop, \emph{BioSystems} 90:295--308

\bibitem{ascolani-lio} Ascolani G and Li\`o P (2014) Modeling $\tgfb$ in early stages of cancer tissue dynamics, \emph{PLOS ONE} 9:e88533

\bibitem{graham-ayati-etal} Graph JM, Ayati BP, Holstein SA, Martin JA (2013) The role of osteocytes in targeted bone remodeling: A mathematical model, \emph{PLoS ONE}:0063884

\bibitem{mullender-huiskes} Mullender MG and Huiskes R (1995) Proposal for the regulatory mechanism of Wolff's law, \emph{J Orthop Res} 13:503--512

\bibitem{vanOers-etal-2008} van Oers RFM, Ruimerman R, Tanck E, Hilbers PAJ, Huiskes R (2008) A unified theory for osteonal and hemi-osteonal remodeling. \emph{Bone} \textbf{42}:250–259

\bibitem{vanOers-etal-2011} van Oers FRM, van Rietbergen B, Ito K, Hilbers PAJ, Huiskes R (2011) A sclerostin-based theory for strain-induced bone formation, \emph{Biomech Model Mechanobiol} 10:663--670

\bibitem{baiotto-etal} Baiotto S, Labat B, Vico L, Zidi M (2009) Bone remodeling regulation under unloading conditions: Numerical investigations, \emph{Comp Biol Med} 39:46--52

\bibitem{adachi-etal-2010} Adachi T, Kameo Y, Hojo M (2010) Trabecular bone remodelling simulation considering osteocytic response to fluid-induced shear stress, \emph{Phil Trans R Soc A} 368:2669--2682

\bibitem{kwon-etal} Kwon J, Naito H, Matsumoto T, Tanaka M (2010) Simulation model of trabecular bone remodeling considering effects of osteocyte apoptosis and targeted remodeling, \emph{J Biomech Science  Eng} 5:539--551

\bibitem{wang-vanoers-etal} Wang H, Ji B, Liu XS, van Oers RFM, Guo XE, Huang Y, Hwang K-C (2014) Osteocyte-viability-based simulations of trabecular bone loss and recovery in disuse and reloading, \emph{Biomech Model Mechanobiol} 13:153--166

\bibitem{andreaus-colloca-iacoviello} Andreaus U, Colloca M, Iacoviello D (2014) Optimal bone density distributions: Numerical analysis of the osteocyte spatial influence in bone remodeling, \emph{Comp Meth Prog Biomed} 113:80--91

\bibitem{pivonka-buenzli-etal-specific-surface} Pivonka P, Buenzli PR, Scheiner S, Hellmich Ch, Dunstan CR (2013) The influence of bone surface availability in bone remodelling---A mathematical model including coupled geometrical and biomechanical regulations of bone cells. \emph{Eng Struct} \textbf{47}:134--147

\bibitem{garciaaznar-rueberg-doblare} Garc\'ia-Aznar JM, Rueberg T, and Doblar\'e M (2005) A bone remodelling model coupling microdamage growth and repair by 3D BMU-activity, \emph{Biomech Model Mechanobiol} 4:147--167

\bibitem{stein-werner} Stein KWH, Werner J (2013) Preliminary analysis of osteocyte lacunar density in long bones of tetrapods: All measures are bigger in sauropod dinosaurs, \emph{PLoS ONE} 8:e77109

\bibitem{lemaire-etal}
Lemaire V \etal\ (2004) Modeling the interactions between osteoblast and osteoclast activities in bone remodelling, \emph{J Theor Biol} 29:293--309

\bibitem{pivonka-etal1}
Pivonka P, Zimak J, Smith DW, Gardiner BS, Dunstan CR, Sims NA, Martin TJ, Mundy GR (2008) Model structure and control of bone remodeling: A theoretical study, \emph{Bone} 43:249

\bibitem{buenzli-etal-anabolic} Buenzli PR, Pivonka P, Gardiner BS, Smith DW (2012) Modelling the anabolic response of bone using a cell population model, \emph{J Theor Biol} \textbf{307}:42--52

\bibitem{buenzli-etal-trabecularisation} Buenzli PR, Thomas CDL, Clement JG, Pivonka P (2013) Endocortical bone loss in osteoporosis: The role of bone surface availability, \emph{Int J Numer Meth Biomed Eng} 29:1307--1322

\bibitem{buenzli-etal-moving-bmu}
Buenzli PR, Pivonka P, Smith DW (2011), Spatio-temporal dynamics of cell distribution in bone multicellular units, \emph{Bone} 48:918

\bibitem{evans-morriss}
Evans DJ and Morriss G (2008) \emph{Statistical Mechanics of Nonequilibrium Liquids}, $2^\text{nd}$ Ed. (Cambridge University Press, Cambridge)

\bibitem{jones-distrib}
Jones DS (1982) \emph{Theory of generalised functions}, $2^\text{nd}$ Ed. (Cambridge University Press, Cambridge)

\bibitem{marotti-zallone-ledda}
Marotti G, Zambonin Zallone A, Ledda M (1976) Number, size and arrangement of osteoblasts in osteons at different stages of formation. \emph{Calcif Tissue Int} 21(suppl):96

\bibitem{jones-ob} Jones SH (1974) Secretory territories and rates of matrix production of osteoblasts, \emph{Calc Tiss Res} \textbf{14}:309--315

\bibitem{pazzaglia-etal-2014} Pazzaglia UE, Congiu T, Sibilia V, Quacci D (2014) Osteoblast-osteocyte transformation. A SEM densitometric analysis of endosteal apposition in rabbit femur, \emph{J Anat} 224:132--141

\bibitem{mader-mueller-etal} Mader KS, Schneider Ph, M\"uller R, Stampanoni M (2013) A quantitative framework for the 3D characterization of the osteocyte lacunar system, \emph{Bone} 57:142--154

\bibitem{dong-peyrin-etal} Dong P, Haupert S, Hesse B, Langer M, Gouttenoire P-J, Bousson V, Peyrin F (2014) 3D osteocyte lacunar morphometric properties and distributions in human femoral cortical bone using synchrotron radiation micro-CT images, \emph{Bone} 60:172--185

\bibitem{vankampen} van Kampen NG (2007) \emph{Stochastic processes in physics and chemistry}. $3^\text{rd}$ Ed (Elsevier)

\bibitem{rumpler-fratzl-etal} Rumpler M, Woesz A, Dunlop J W C, van Dongen J T and Fratzl P (2008) The effect of geometry on three-dimensional tissue growth. \textit{J. R. Soc. Interface} \textbf{5}:1173--1180

\bibitem{bidan-fratzl-dunlop-etal-plos1} Bidan CM, Kommareddy KP, Rumpler M, Kollmannsberger Ph, Br\'echet YJM, Fratzl P, Dunlop JWC (2012) How linear tension converts to curvature: Geometric control of bone tissue growth, \emph{PLoS ONE} 7:e36336

\bibitem{bidan-fratzl-dunlop-etal-bis} Bidan CM, Kommareddy KP, Rumpler M, Kollmannsberger Ph, Fratzl P, Dunlop JWC (2013) Geometry as a factor for tissue growth: Towards shape optimization of tissue engineering scaffolds, \emph{Adv Healthcare Mater} 2:186--194

\bibitem{pazzaglia-etal-2010} Pazzaglia UE, Congiu T, Marchese M,  Dell'Orbo C (2010) The shape modulation of osteoblast--osteocyte transformation and its correlation with the fibrillar organization in secondary osteons, \emph{Cell Tissue Res} \textbf{340}:533--540


\bibitem{qiu-parfitt-etal-2010} Qiu S, Rao DS, Palnitkar S, Parfitt AM (2010) Dependence of bone yield (volume of bone formed per unit of cement surface area) on resorption cavity size during osteonal remodeling in human rib: Implications for osteoblast function and the pathogenesis of age-related bone loss, \emph{J Bone Miner Res} \textbf{25}:423--430


\bibitem{lim-etal} Lim JY, Shaughnessy MC, Zhou Z, Noh H, Vogler EA, Donahue HJ (2008) Surface energy effects on osteoblast spatial growth and mineralization, \emph{Biomaterials} 29:1776--1784

\bibitem{metz-martin-turner}
Metz LN, Martin RB, and Turner AS (2003) Histomorphometric analysis of the effects of osteocyte density on osteonal morphology and remodeling, \emph{Bone} 33:753

\bibitem{sethian} Sethian JA (1999) \emph{Level set methods and fast marching methods} (Cambridge University Press)

\bibitem{osher-fedkiw} Osher S, Fedkiw R (2003) \emph{Level set methods and dynamic implicit surfaces} (New York: Springer)

\bibitem{vashishth-fyhrie-etal} Vashishth D, Gibson G, Kimura J, Schaffler MB, Fyhrie DP (2002) Determination of bone volume by osteocyte population, \emph{Anat Rec} 267:292--295

\bibitem{qiu-parfitt-etal-abstract} Qiu S, Palnitkar S, Rao D, Parfitt AM (2000) Is osteocyte density determined by osteoblasts in bone remodelling? \emph{ASBMR 22nd Annual Meeting}, SA008, p.~S236

\bibitem{zarrinkalam-atkins-etal} Zarrinkalam MR, Mulaibrahimovic A, Atkins GJ, Moore RJ (2012) Changes in osteocyte density correspond with changes in osteoblast and osteoclast activity in an osteoporotic sheep model, \emph{Osteoporos Int} 23:1329--1336

\bibitem{pazzaglia-etal-2012} Pazzaglia UE, Congiu T, Franzetti E, Marchese M, Spagnuolo F, Di Mascio L, Zarattini G (2012) A model of osteoblast-osteocyte kinetics in the development of secondary osteons in rabbits, \emph{J Anat} 220:372--383

\bibitem{hernandez-majeska-schaffler} Hernandez CJ, Majeska RJ, Schaffler MB (2004) Osteocyte density in woven bone, \emph{Bone} 35:1095--1099

\bibitem{qiu-parfitt-etal-2002} Qiu S, Rao DS Palnitkar S, Parfitt AM (2002) Relationship between osteocyte density and bone formation rate in human cancellous bone, \emph{Bone} 31:709--711

\bibitem{qiu-parfitt-etal-2002b} Qiu S, Rao DS Palnitkar S, Parfitt AM (2002) Age and distance from the surface, but not menopause reduce osteocyte density in human cancellous bone, \emph{Bone} 31:313--318

\bibitem{sharma-etal} Sharma D, Ciani C, Ramirez Marin PA, Levy JD, Doty SB, and Fritton SP (2012) Alterations in the Osteocyte Lacunar-Canalicular Microenvironment due to Estrogen Deficiency, \emph{Bone} 51:488--497

\bibitem{carter-thomas-clement-etal} Carter Y, Thomas CDL, Clement JG, Peele AG, Hannah K, Cooper DML (2013) Variation in osteocyte lacunar morphology and density in the human femur---a synchrotron radiation micro-CT study \emph{Bone} 52:126--132

\bibitem{mullender-huiskes-etal-1996} Mullender MG, Huiskes R, Versleyen H, Buma P (1996) Osteocyte density and histomorphometric parameters in cancellous bone of the proximal femur in five mammalian species, \emph{J Orhop Res} 14:972--979

\bibitem{parfitt-1994} Parfitt A M (1994) Osteonal and hemi-osteonal remodeling: the spatial and temporal framework for signal traffic in adult human bone. \textit{J. Cell. Biochem.} \textbf{55}:273--286

\bibitem{sims-martin-2014} Sims NA and Martin TJ (2014) Coupling the activities of bone formation and resorption: A multitude of signals within the basic multicellular unit, \emph{BoneKEy Reports} 3:481

\bibitem{cacchioli-etal-MAR-rabbit} Cacchioli A, Ravanetti F, Soliani L, Borhetti P (2012) Preliminary study on the mineral apposition rate in distal femoral epiphysis of New Zealand white rabbit at skeletal maturity, \emph{Anat Histol Embryol} 41:163--169

\bibitem{adachi-etal-2009} Adachi T, Aonuma Y, Taira K, Hojo M, Kamioka H (2009) Asymmetric intercellular communication between bone cells: propagation of the calcium signaling. \emph{Biochem Biophys Res Commun} 389:495--500

\bibitem{jilka-etal} Jilka RL, Noble B, Weinstein RS (2013) Osteocyte apoptosis, \emph{Bone} 54:264

\bibitem{kamioka-etal} Kamioka H, Honjo T, Takano-Yamamoto T (2001) A Three-dimensional distribution of osteocyte processes revealed by the combination of confocal laser scanning microscopy and differential interference contrast microscopy, \emph{Bone} 28:145--149

\bibitem{lange} Lange R-J (2012) Potential theory, path integrals and the Laplacian of the indicator, \emph{J High Energy Phys} 2012:32

\end{thebibliography}


\end{document}